\documentclass[11pt]{article}

\usepackage{amsmath,amsfonts,amssymb,epsfig,hyperref,a4}

\renewcommand\slash[1]{\not \! #1}
\newcommand{\ud}{\mathrm{d}}
\providecommand{\sigmadip}{\hat{\sigma}}
\def\mbf#1{\mathchoice{\hbox{\boldmath $\displaystyle #1$}}
        {\hbox{\boldmath $\textstyle #1$}}{\hbox{\boldmath $\scriptstyle #1$}}
        {\hbox{\boldmath $\scriptscriptstyle #1$}}}

\begin{document}

%
\def\be{\begin{equation}}
\def\ee{\end{equation}}
\def\bea{\begin{eqnarray}}
\def\eea{\end{eqnarray}}
\def\rarr{\rightarrow}
\def\nn{\nonumber}
\def\fr{\frac}
\renewcommand\slash[1]{\not \! #1}
\newcommand\qs{\!\not \! q}
\def\del{\partial}
\def\gam{\gamma}
\newcommand\vphi{\varphi}
\def\tr{\mbox{tr}\,}
\newcommand\nin{\noindent}

\def\del{\partial}
\def\pbar{\bar{p}}
\begin{titlepage}
\begin{flushright}
ZU-TH 02/12, MZ-TH/12-06
\end{flushright}
\vspace{0.6cm}
\begin{center}
\boldmath
{\LARGE{\bf The New $F_L$ Measurement from HERA}}\\[.2cm]
{\LARGE{\bf and the Dipole Model}}\\
\unboldmath
\end{center}
\vspace{0.6cm}
\begin{center}
{\bf \Large
Carlo Ewerz\,$^{a,b,1}$, Andreas von Manteuffel\,$^{c,d,2}$,\\
Otto Nachtmann\,$^{a,3}$, Andr\'e Sch\"oning\,$^{e,4}$}
\end{center}
\vspace{.2cm}
\begin{center}
$^a$
{\sl
Institut f\"ur Theoretische Physik, Universit\"at Heidelberg\\
Philosophenweg 16, D-69120 Heidelberg, Germany}
\\[.5cm]
$^b$
{\sl
ExtreMe Matter Institute EMMI, GSI Helmholtzzentrum f\"ur Schwerionenforschung\\
Planckstra{\ss}e 1, D-64291 Darmstadt, Germany}
\\[.5cm]
$^c$
{\sl
Institut f\"ur Theoretische Physik, Universit\"at Z\"urich\\
Winterthurerstr.\ 190, CH-8057 Z\"urich, Switzerland
}
\\[.5cm]
$^d$
{\sl
Institut f\"ur Physik (THEP), Johannes-Gutenberg-Universit\"at\\
D-55099 Mainz, Germany}
\\[.5cm]
$^e$
{\sl
{\sl
Physikalisches Institut, Universit\"at Heidelberg\\
Philosophenweg 12, D-69120 Heidelberg, Germany}
}
\end{center}                                                                
\vfill
\begin{abstract}
\noindent
From the new measurement of $F_L$ at HERA we derive fixed-$Q^2$ 
averages $\langle F_L / F_2 \rangle$. We compare these with bounds 
which are rigorous in the framework of the standard dipole picture. 
The bounds are sharpened by including information on the charm 
structure function $F_2^{(c)}$. Within the experimental errors the bounds 
are respected by the data. But for $3.5\,\mbox{GeV}^2 \le Q^2 \le 20\,\mbox{GeV}^2$ 
the central values of the data are close to and in some cases even 
above the bounds. Data on $F_L/F_2$ significantly exceeding the bounds 
would rule out the standard dipole picture at these kinematic points. 
We discuss, furthermore, how data respecting 
the bounds but coming close to them can give information on questions 
like colour transparency, saturation and the dependencies of the dipole-proton 
cross section on the energy and the dipole size. 
\vfill
\end{abstract}
\vspace{5em}
\hrule width 5.cm
\vspace*{.5em}
{\small \noindent
$^1$ email: C.Ewerz@thphys.uni-heidelberg.de\\
$^2$ email: manteuffel@uni-mainz.de\\
$^3$ email: O.Nachtmann@thphys.uni-heidelberg.de\\
$^4$ email: schoening@physi.uni-heidelberg.de
}
\end{titlepage}

\section{Introduction}
\label{sec:intro}

Recently new results for the structure functions $F_L$ and $F_2$ of 
deep inelastic electron- and positron-proton scattering (DIS) have 
been published by the H1 Collaboration \cite{Collaboration:2010ry}. 
In this note we compare these results with predictions of the popular 
colour-dipole model of DIS. That is, we investigate if the data respect 
certain bounds for the ratios of structure functions. These bounds are 
rigorous predictions of the dipole model and rely only on the 
non-negativity of the dipole-proton cross section. 

The kinematics of $e^\pm p$ scattering is well known, see for 
instance \cite{Collaboration:2010ry,Nachtmann:1990ta}. 
The reaction is 
\be\label{1.1}
e^\pm (k)  + p(p) \longrightarrow e^\pm (k') + X(p') 
\ee
and we use the variables 
\be\label{1.2}
\begin{aligned}
q &= k - k' = p' - p\,, \:\:\:\:\:\:\:\:\:\:\:
&Q^2 &= -q^2\,,\\
W^2 &= (p + q)^2 \,,
&x &= \frac{Q^2}{2 pq} = \frac{Q^2}{W^2 + Q^2 - m_p^2} \,.
\end{aligned}
\ee
The measured structure functions $F_2$ and $F_L$ are related to 
the cross sections $\sigma_T$ and $\sigma_L$ for absorption of 
transversely or longitudinally polarised virtual photons by
\be\label{1.3}
\begin{split}
F_2(x, Q^2) &=
        \frac{Q^2}{4 \pi^2 \alpha_{\rm em}} \, (1-x) 
                \left[ \sigma_T(x,Q^2) + \sigma_L(x,Q^2) \right] \,,
\\
F_L(x, Q^2) &=
        \frac{Q^2}{4 \pi^2 \alpha_{\rm em}} \, (1-x) \, \sigma_L(x,Q^2) \,.
\end{split}
\ee
Here Hand's convention \cite{Hand:1963bb} 
for the virtual-photon flux factor is 
used and terms of order $m^2_p/W^2$ are neglected. For low to 
moderate values of $Q^2$ the dipole picture for DIS 
\cite{Nikolaev:1990ja,Nikolaev:et,Mueller:1993rr} 
is frequently used to describe the data. For various applications of 
the dipole model see for instance 
\cite{Golec-Biernat:1998js}-\cite{Albacete:2009fh}. 
In \cite{Ewerz:2004vf,Ewerz:2006vd} this dipole picture was 
thoroughly examined using functional methods 
of quantum field theory. In particular, the assumptions were spelled 
out which one has to make in order to arrive at the standard dipole-model 
formulae for $\sigma_T$ and $\sigma_L$ or, equivalently, $F_2$ and $F_L$, 
\be
\begin{split}
\label{1.4}
F_2 (x,Q^2)&= \frac{Q^2}{4 \pi^2 \alpha_{\rm em}} \, (1-x) 
\sum_q\int \ud^2 r \,
\left[ w^{(q)}_T(r,Q^2) + w^{(q)}_L(r,Q^2) \right] 
\sigmadip^{(q)}(r,\xi) \,,
\\
F_L(x,Q^2) &= 
\frac{Q^2}{4 \pi^2 \alpha_{\rm em}} \, (1-x) 
\sum_q\int \ud^2 r \, w^{(q)}_L(r,Q^2)\,
\sigmadip^{(q)}(r,\xi) \,,
\end{split}
\ee
see section 6 of \cite{Ewerz:2006vd}. In \eqref{1.4} $w^{(q)}_{T,L}$ are the 
probability densities for the virtual photon $\gamma^*$ splitting into a 
quark-antiquark pair of flavour $q$ and transverse separation $r$. 
Their standard expressions are given in Appendix \ref{app1}.  
An integration over the quark's longitudinal momentum is performed. 
The cross section for the $q\bar q$ pair scattering on the proton is 
denoted by $\hat\sigma^{(q)}(r,\xi)$. This cross section depends 
on $r$ and an energy variable $\xi$ the choice of which is left open here. 
In \cite{Ewerz:2004vf,Ewerz:2006vd,Ewerz:2006an,Ewerz:2007md,Ewerz:2011ph} 
it was argued that the correct variable to choose is $\xi=W$. 
However, in the literature the energy variable used most frequently in 
the dipole cross section is $\xi=x$. 

In the standard dipole model formulae \eqref{1.4} the densities $w^{(q)}_{T,L}$ 
are known (see Appendix \ref{app1}) but the 
dipole-proton cross sections $\hat\sigma^{(q)}$ have to be taken from 
a model. In the following we shall only use that they have to be non-negative, 
\be\label{1.5}
\hat\sigma^{(q)}(r,\xi)\geq 0 \,.
\ee
This alone allows to derive a non-trivial upper bound, valid in any 
dipole model, on the ratio
\be\label{1.6}
R(x,Q^2) = \frac{\sigma_L (x,Q^2)}{\sigma_T (x,Q^2)}
\ee
see \cite{Ewerz:2006vd,Ewerz:2006an}. Equivalently, one can obtain a 
non-trivial upper bound on the ratio 
\be\label{1.6a}
\frac{F_L (x,Q^2)}{F_2 (x,Q^2)} = \frac{R(x,Q^2)}{1+R(x,Q^2)} \,.
\ee
This bound can be substantially 
improved if information on the charm structure function $F^{(c)}_2(x,Q^2)$ 
is included \cite{Ewerz:2007md}. 
There is then an allowed domain, again valid in any dipole model, for 
the two-dimensional vector
\be\label{1.7}
\vec{V}(x,Q^2)= 
\begin{pmatrix} F_L(x,Q^2)/F_2(x,Q^2) \\
                F_2^{(c)}(x,Q^2)/F_2(x,Q^2) \end{pmatrix} 
\,.
\ee

It is the purpose of this note to confront the dipole-model bounds on 
$F_L/F_2$ and on the vector $\vec V(x,Q^2)$ with the new HERA 
results \cite{Collaboration:2010ry}. This is done in section \ref{sec:2}. 
In section \ref{sec:3} we discuss the results, and we give a summary in section \ref{sec:4}. 

\section{The dipole-model bounds and the data}
\label{sec:2}

\begin{figure}
\hspace*{-1.5cm}
\includegraphics[height=6.3cm]{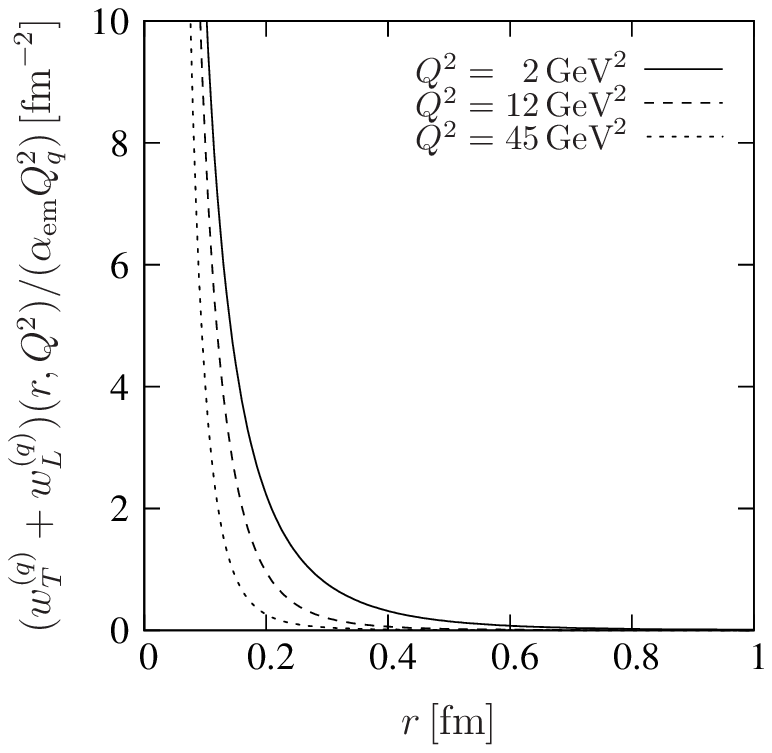}
\includegraphics[height=6.3cm]{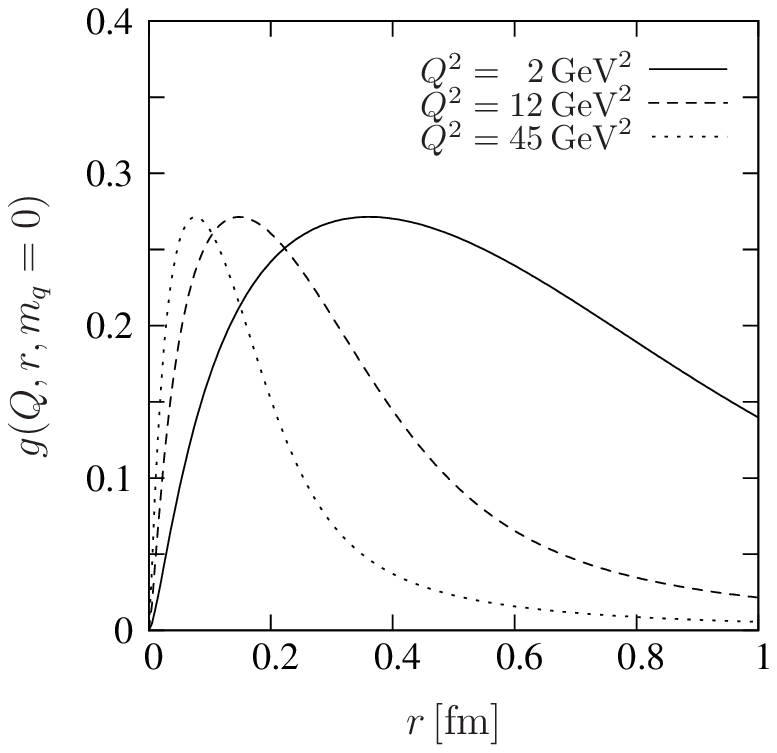}
\caption{The functions $\frac{1}{\alpha_{\rm em} Q_q^2}(w_T^{(q)} + w_L^{(q)})(r,Q^2)$ (left) 
and $g(Q,r,m_q)$ (right) versus $r$, both for three fixed values of $Q^2$ and for quark 
mass $m_q=0$\,; see \eqref{1.4} and \eqref{2.1}.
\label{fig1}}
\end{figure}

We discuss first the bound for the ratio $F_L/F_2$ of \eqref{1.6a}. For this we define
\be\label{2.1}
g(Q,r,m_q) = \frac{w_L^{(q)}(r,Q^2)}{w_T^{(q)}(r,Q^2) + w_L^{(q)}(r,Q^2)} \,,
\ee
where $m_q$ is the mass of the quark $q$. 
For the case of massless quarks, $m_q=0$, figure \ref{fig1} shows 
$\frac{1}{\alpha_{\rm em} Q_q^2}(w_T^{(q)} + w_L^{(q)})(r,Q^2)$ and $g(Q,r,0)$ as functions of $r$ 
for three different values of $Q=\sqrt{Q^2}$ (compare figure 10 
of \cite{Ewerz:2006vd} for a similar plot of the function $(w_L^{(q)}/w_T^{(q)})(r,Q^2)$). 
Note that $(w_T^{(q)} + w_L^{(q)})(r,Q^2)$ is monotonously decreasing 
with $r$. Its behaviour for small and large $r$ is as follows for $m_q=0$: 
\be
\label{9a}
\begin{split}
&(w_T^{(q)} + w_L^{(q)})(r,Q^2)  \propto \frac{1}{r^2} \:\:\:\:\:\: \mbox{for}\:\:\:\:\:\: r \to 0 \,,
\\
&(w_T^{(q)} + w_L^{(q)})(r,Q^2)  \propto \frac{1}{r^4} \:\:\:\:\:\: \mbox{for}\:\:\:\:\:\: r \to \infty \,.
\end{split}
\ee
For a derivation of these results and for the case $m_q \neq 0$ see 
appendix A of \cite{Ewerz:2011ph}. 
For massless quarks the function $g$ depends only on the dimensionless 
variable 
\be\label{2.2}
z=Qr \,, 
\ee
such that we can write 
\be
\label{2.2a}
\tilde{g} (z) = g(Q,r,0) \,.
\ee
The function $\tilde{g}(z)$ has a maximum at 
\be\label{2.3}
z_m = 2.5915
\ee
with
\be\label{2.4}
\tilde{g}(z_m)=0.27139\,.
\ee
It was shown in \cite{Ewerz:2007md} that 
\be\label{2.5}
g(Q,r,m_q) \le \tilde{g} (z_m) 
\ee
for all $Q \geq 0$, $r\geq 0$ and $m_q \geq 0$. Using then \eqref{1.5} the dipole-model 
formulae \eqref{1.4} lead to the bound 
\be\label{2.6}
\frac{F_L(x,Q^2)}{F_2(x,Q^2)} \le \tilde{g}(z_m) = 0.27139 \,. 
\ee
We note that the bound \eqref{2.6} for $F_L/F_2$ is equivalent to the 
bound for $R$ \eqref{1.6} derived in \cite{Ewerz:2006an,Ewerz:2007md}, 
\be
\label{15new}
R(x,Q^2) \le \frac{\tilde{g}(z_m)}{1-\tilde{g}(z_m)} = 0.37248\,.
\ee

Data for $F_L$ and $F_2$ at the same kinematic points are presented 
in \cite{Collaboration:2010ry} for $Q^2$ values ranging from $1.5$ to 
$45$ $\mbox{GeV}^2$. 
The data for the same $Q^2$ value 
span a small range of $x$ and this range varies strongly with $Q^2$; 
see figure 12 of \cite{Collaboration:2010ry}. 
On the other hand, for all $Q^2$ bins the data are inside a narrow
$W$ interval 
\be\label{2.6a} 
167 \,\mbox{GeV} - 232 \,\mbox{GeV} 
\ee
with a mean value of about $W_0=200$ GeV. 
Therefore, in the following we find it more convenient to consider $F_L$ 
and $F_2$ as functions of $W$ and $Q^2$ instead of $x$ and $Q^2$. 

\begin{figure}
\begin{center}
\includegraphics[width=0.8\textwidth]{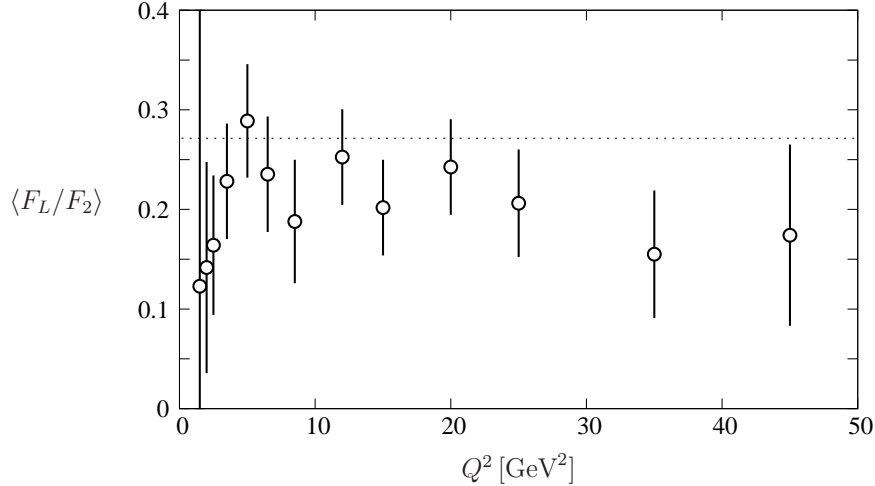}
\end{center}
\caption{The data for the fixed-$Q^2$ averages $\langle F_L/F_2 \rangle$ 
confronted with the dipole-model upper bound \eqref{2.6} represented by the 
dotted line. The data are extracted from \cite{Collaboration:2010ry}. 
\label{fig2new}}
\end{figure}

Since we do not expect any large variation of the ratio $F_L(W,Q^2) / F_2(W,Q^2)$ 
for fixed $Q^2$ within the $W$ interval \eqref{2.6a} of the 
measurement we have averaged the H1 data \cite{Collaboration:2010ry} 
for given $Q^2$. 
Error weighted averages $\langle F_L(W,Q^2)/F_2(W,Q^2) \rangle$ are calculated taking into
account the total uncorrelated and correlated experimental uncertainties. 
The averages are confronted with the bound \eqref{2.6} in figure \ref{fig2new}. 

We note firstly, that electromagnetic gauge invariance requires
\be\label{2.7}
\frac{F_L(W,Q^2)}{F_2(W,Q^2)} \, \longrightarrow \, 0 
\ee
for $Q^2\to 0$ at fixed $W$. The data indicate, indeed, a decrease 
of $F_L/F_2$ for small $Q^2$. Fitting $F_L/F_2$ with a constant value, as done 
in \cite{Collaboration:2010ry}, does not seem very plausible 
physically, in view of \eqref{2.7}.

The second point to note is that the data in figure \ref{fig2new} are rather 
close to the upper bound \eqref{2.6} from the dipole model, especially so 
for 
\be\label{2.8}
3.5 \,\mbox{GeV}^2 \le Q^2 \le 20 \,\mbox{GeV}^2 \,.
\ee

The bound \eqref{2.6} on $F_L/F_2$ can be improved if one takes into 
account that there is a non-vanishing contribution from charm quarks 
to $F_L$ and $F_2$, see \cite{Ewerz:2007md}. Specifically, 
considering massless $u,d$ and $s$ 
quarks, a massive $c$ quark and neglecting $b$ 
quarks we can derive certain allowed domains for the vector 
$\vec V(x,Q^2)$ \eqref{1.7} from the dipole model. Again these domains 
depend only on the known photon densities $w^{(q)}_{T,L}$, see 
\eqref{denst}-\eqref{sumpsiLdens}, and on the non-negativity 
of the cross sections $\hat\sigma^{(q)}$, see \eqref{1.5}. That is, for any 
dipole model with the standard photon probability densities 
$w^{(q)}_{T,L}$ the vector $\vec V(x,Q^2)$ must be inside the appropriate allowed 
domain for the given $Q^2$ value. A detailed description of how these 
domains are obtained has been given in \cite{Ewerz:2007md}. 
The allowed domains can be understood as correlated bounds 
for the ratios $F_L/F_2$ and $F_2^{(c)}/F_2$. More precisely, one 
obtains for any given $x$ and $Q^2$ an upper bound on $F_L/F_2$ 
which depends on the value of $F_2^{(c)}/F_2$ at the same kinematic point. 

\begin{figure}
\includegraphics[width=\textwidth]{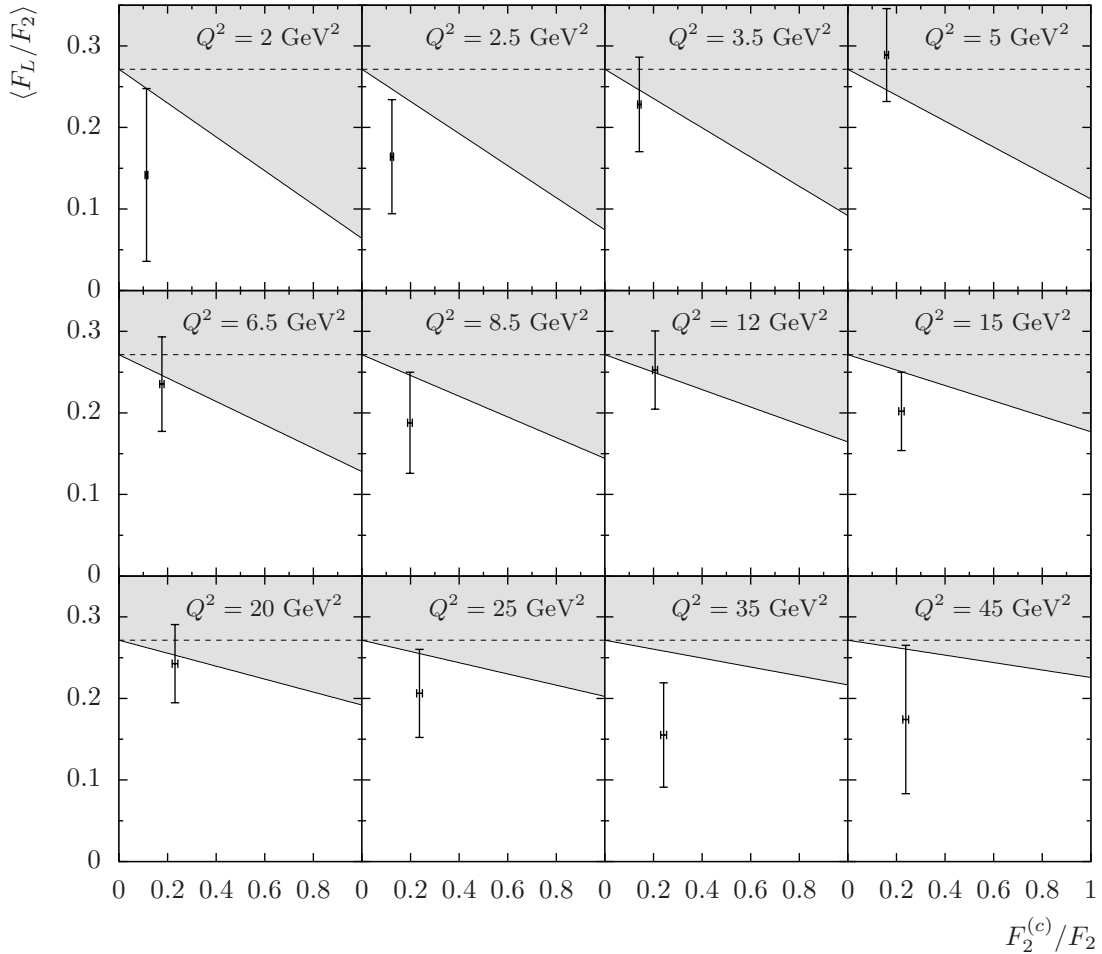}
\caption{The allowed domains and the data for $\langle F_L/F_2 \rangle$ versus 
$F_2^{(c)}/F_2$ for $Q^2 = 2\,\mbox{GeV}^2$ to $Q^2 = 45\,\mbox{GeV}^2$. 
In the dipole model, the shaded areas are excluded by the 
correlated bounds for $F_L/F_2$ and $F_2^{(c)}/F_2$. The dotted line is the 
bound \eqref{2.6} on $F_L/F_2$ only. 
\label{fig4}}
\end{figure}

In figure \ref{fig4} we show these allowed domains and the corresponding data. 
The bounds are calculated for a charm quark mass of $m_c=1.23$ GeV. 
For each data point the corresponding ratio $F_2^{(c)}/F_2$ is obtained 
using NLO QCD calculations provided by the OPENQCDRAD package \cite{OPENQCDRAD}, 
again with a charm pole mass of $m_c = 1.23$ GeV.
For this calculation the JR09FFNNLO parametrisation \cite{JimenezDelgado:2008hf} 
of the proton parton density functions was used, which was found to describe 
preliminary HERA charm data \cite{Corradi:2010zz} very well 
within the experimental correlated uncertainties of typically 3-9\%.
Here and in the following we do not consider the data point at $Q^2=1.5\,\mbox{GeV}^2$ 
from \cite{Collaboration:2010ry} as it has an exceedingly large error. 

\begin{figure}
\begin{center}
\includegraphics[width=0.85\textwidth]{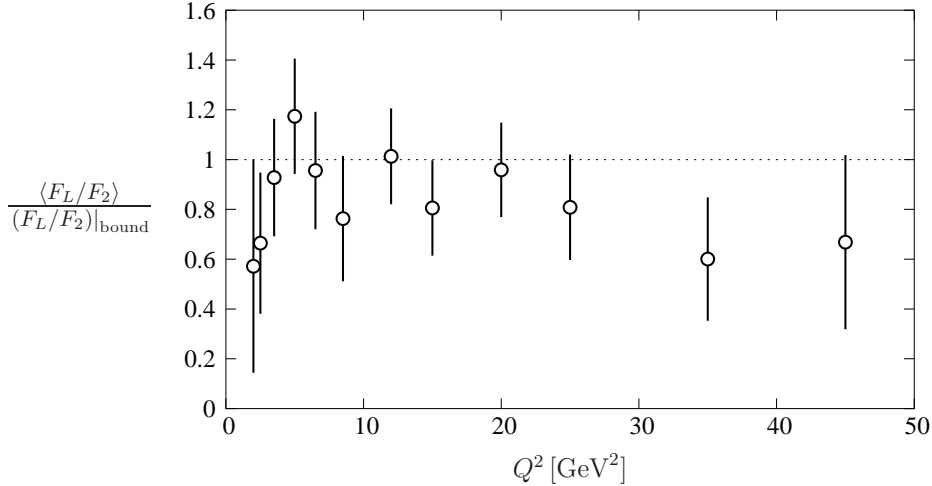}
\end{center}
\caption{The ratio $\langle F_L/F_2 \rangle/\left.(F_L/F_2) \right|_{\rm bound}$, 
see \eqref{2.51}, where the bound on $F_L/F_2$ results from taking into 
account the value of $F_2^{(c)}/F_2$ at the kinematical point of each data point. 
\label{fig5}}
\end{figure}

The significance of the data points in relation to the bound can be 
seen more clearly from the quantity
\be
\label{2.51}
\frac{\langle F_L/F_2 \rangle}{\left. (F_L/F_2) \right|_{\rm bound}}
\ee
which we plot in figure \ref{fig5}. 
For this figure the bound $\left. (F_L/F_2) \right|_{\rm bound}$ 
for each data point is extracted from figure \ref{fig4} taking into 
account the corresponding value of $F_2^{(c)}/F_2$ at that kinematical 
point. 

\section{Discussion}
\label{sec:3}

We see from figures \ref{fig2new}-\ref{fig5} that the data 
for $\langle F_L/F_2 \rangle$ as derived from \cite{Collaboration:2010ry} 
come very close to the bounds which result from the dipole picture. 
We now discuss the meaning of this observation from the points of view 
of both, a dipole-model enthusiast, and a dipole-model sceptic, 
respectively. 

\subsection*{Dipole-model enthusiast's view}

The dipole-model enthusiast will say that within the errors of the data 
the bounds are respected. Furthermore, he can use the data to give 
qualitative arguments concerning the behaviour of the dipole-proton 
cross sections for small and large radii $r$. Let us assume power 
behaviour of $\hat\sigma^{(q)}(r,\xi)$ for $r \to 0$ and $r \to \infty$, 
\be
\label{100}
\begin{split}
 \hat\sigma^{(q)}(r,\xi) \propto r^a \:\:\:\:\:\: &\mbox{for}\:\:\:\:\:\: r \to 0 \,,
\\
 \hat\sigma^{(q)}(r,\xi) \propto r^{2-b}\:\:\:\:\:\: &\mbox{for}\:\:\:\:\:\: r \to \infty \,.
\end{split}
\ee
Taking into account \eqref{9a} we find that the integrals for $F_2$ and $F_L$ 
in \eqref{1.4} are convergent if 
\be
\label{101}
a>0 \:\:\:\:\:\: \mbox{and} \:\:\:\:\:\:  b>0\,.
\ee
Of course, with the usual assumptions of colour transparency for small $r$, 
implying $a=2$, and of saturation for the dipole-proton cross sections 
for large $r$, implying $b=2$, the requirements \eqref{101} are satisfied. 
From the experimental findings of figures \ref{fig2new} to \ref{fig5} we 
can now give qualitative arguments based on the data, that the exponents 
$a$ and $b$ in \eqref{100} cannot be too small. Indeed, for a small value of $a$ 
the cross sections $\hat\sigma^{(q)}(r,\xi)$ would decrease only slowly for 
$r \to 0$ and this region of small $r$ would contribute significantly in the 
integrals \eqref{1.4}. But, as we see from the second plot in figure \ref{fig1}, 
the function $g(Q,r,0)$ is small there and this would lead to a small value 
for $F_L/F_2$, much below the bound \eqref{2.6}, contrary to what is 
seen in the data. A similar argument applies to the exponent $b$ in 
\eqref{101}, considering the large $r$ behaviour of $g(Q,r,0)$ in figure 
\ref{fig1}. Thus, the dipole-model enthusiast may hope that with more 
data it may even be possible to determine the exponents $a$ and $b$ 
from the data on $F_L/F_2$ directly without making model 
assumptions for $\hat\sigma^{(q)}(r,\xi)$. 

\subsection*{Dipole-model sceptic's view} 

Let us now go over to the point of view of the dipole-model sceptic. He will 
note that some central values of the data for $F_L/F_2$ in figure \ref{fig4} 
are, in fact, {\sl above} the corresponding bound. If any of the measured points with 
$\langle F_L/F_2 \rangle > \left. (F_L/F_2) \right|_{\rm bound}$ 
is confirmed, with corresponding small error, 
by further experiments then, as a clear consequence, the standard 
dipole picture would not be valid at this kinematic point. But what 
would be the consequences if the bound for $F_L/F_2$ is not {\em violated} 
but {\em saturated}\,? 

For the sake of the argument we shall now for a moment {\em assume} that the 
bound for $F_L/F_2$ is reached in the $Q^2$ range \eqref{2.8}. 
Clearly, this is not incompatible with the data, see figures \ref{fig4} and \ref{fig5}. 
The consequence is that the dipole-proton cross 
sections $\hat\sigma^{(q)}(r,\xi)$ in \eqref{1.4} should only contribute 
at that particular $r$ values where the functions $g(Q,r,0)$ and $g(Q,r,m_c)$ 
of \eqref{2.1} have their maximum. This is for both functions the case for 
\be
\label{200}
r \approx \frac{0.51 \,\mbox{fm}}{\sqrt{Q^2/\mbox{GeV}^2}} \,.
\ee
We see this for $g(Q,r,0)$ from the second plot in figure \ref{fig1} and this 
also holds for $g(Q,r,m_c)$. Thus, the cross sections $\hat\sigma^{(q)}(r,\xi)$ 
should be strongly peaked at these $r$ values for the whole $Q^2$ interval 
\eqref{2.8}, something like a $\delta$ function 
\be\label{3.2}
\hat{\sigma}^{(q)} (r,\xi) \, \approx \, \delta \left( r - \frac{0.51}{Q} \right) 
\:\:\:\:\:\:\:\:\:\: \mbox{($r$ in fm, $Q$ in GeV)}
\,.
\ee
The corresponding $r$ values range from $0.27$ fm for $Q^2= 3.5 \,\mbox{GeV}^2$ 
to $0.11$ fm for $Q^2= 20 \,\mbox{GeV}^2$. 
With increasing $Q$ the position of the delta function peak 
in \eqref{3.2} moves to smaller $r$ values. As we have argued at 
length in \cite{Ewerz:2004vf,Ewerz:2006vd,Ewerz:2011ph}, 
the correct energy variable in the dipole-proton cross section 
$\hat \sigma^{(q)}(r,\xi)$ is $\xi=W$. Since the data on $F_L/F_2$ 
is essentially at one value of $W \approx W_0=200$ GeV 
(more precisely, in the narrow range \eqref{2.6a} around $W_0$) 
we get from \eqref{3.2}
a $Q^2$ dependence in $\hat\sigma^{(q)}(r,W_0)$ which 
should {\em not} be there. The conclusion is that a saturation of the 
bound on $F_L/F_2$ in a whole $Q^2$ interval as in \eqref{2.8} is 
incompatible with the dipole model and the dipole-proton cross sections 
having the correct functional dependence $\hat\sigma^{(q)}(r,W)$.

With the -- incorrect -- choice of energy variable 
$\xi=x$ in $\hat\sigma^{(q)}(r,\xi)$ we get the following. Since 
the data on $F_L/F_2$ is essentially at $W=W_0$ (namely in the narrow 
range \eqref{2.6a} around $W_0$), we have from \eqref{1.2}
\be\label{3.4}
x \simeq \frac{Q^2}{W_0^2} \,, \:\:\:\:\:\: 
Q \simeq \sqrt{x} \,W_0 \,.
\ee
Inserting this in \eqref{3.2} gives 
\be\label{3.5}
\hat{\sigma}^{(q)}(r,x) \, \approx \, \delta\left(r-\frac{0.51}{\sqrt{x} \,W_0}\right) 
\:\:\:\:\:\:\:\:\:\: \mbox{($r$ in fm, $W_0$ in GeV)} \,.
\ee
Thus, there is in this case no immediate conflict with the functional 
dependence $\hat\sigma^{(q)}(r,x)$. But we note that 
as $x$ {\em decreases} the peak of the cross section $\hat\sigma^{(q)}(r,x)$ 
in \eqref{3.2} shifts to {\em larger} values of $r$. 
This is in contrast to what one finds in popular dipole models, like the one 
invented by Golec-Biernat and W\"usthoff \cite{Golec-Biernat:1998js}. 
There, one assumes  a dipole-proton cross section saturating at 
large $r$ with an $x$-dependent saturation scale. But in that model 
for {\em decreasing} values of $x$ the cross section $\hat\sigma^{(q)}(r,x)$ 
moves to {\em smaller} values of $r$, see figure 2 of \cite{Golec-Biernat:1998js}. 
This is in contradiction to what we found above in \eqref{3.5}. 

The dipole-model sceptic could, furthermore, argue as follows. Since the 
bounds explored in the present paper are just more or less satisfied by 
the data it will certainly pay to explore further rigorous bounds which can be 
constructed using the methods of \cite{Ewerz:2007md}. One could, for 
instance, consider correlated bounds on $F_L/F_2$ at different $Q^2$ values. 
It remains to be explored if the dipole model survives such extended tests. 

\section{Summary}
\label{sec:4}

In this paper we have compared the recent data on $F_L/F_2$ -- to be 
precise: their fixed-$Q^2$ averages -- with rigorous bounds derived 
in the framework of the dipole model. Within the experimental errors 
the bounds are satisfied. But the data is surprisingly close to the bounds 
for $3.5 \,\mbox{GeV}^2 \le Q^2 \le 20 \,\mbox{GeV}^2$. We have discussed 
the meaning of these findings from the points of view of both, the dipole-model 
enthusiast and the sceptic. The enthusiast will have to admit that the sceptic's 
arguments could give problems to the dipole picture if the central values of 
the data are confirmed with small errors by further experiments. The sceptic 
will have to concede that, given the errors of the data, $\delta$ functions 
for the cross sections $\hat\sigma^{(q)}(r,\xi)$ as in \eqref{3.2} and \eqref{3.5} 
are not really necessary and that the widths of the distributions 
compatible with the data have to be explored. Thus, given the present data, 
we must leave it 
to the reader if he will join the camp of the enthusiast or that of the sceptic. 
More data with small errors would be needed to decide the issue. 
In any case we hope to have demonstrated in our paper that measurements 
of $F_L/F_2$ give very valuable information on the dipole picture, its 
validity, and potentially 
on questions like colour transparency and saturation of the dipole-proton 
cross section. Thus, programs for future electron- and positron-proton 
scattering experiments (see for instance \cite{eicwhitepaper}, \cite{Klein:2010zz}) 
certainly should foresee $F_L$ measurements as an important item on the list of 
physics topics. 

\section*{Acknowledgements}
We are grateful to Sascha Glazov for providing data tables
and information on correlated data uncertainties.
We would like to thank Markus Diehl for useful discussions. 
The work of C.\,E.\ was supported by the Alliance Program of the
Helmholtz Association (HA216/EMMI). The work of 
A.\,v.\,M.\ was supported by the Schweizer Nationalfonds (Grant
200020\_124773/1). 

\begin{appendix}

\section{Photon densities}
\label{app1}

The probability densities $w^{(q)}_{T,L}$ for the virtual photon in 
\eqref{1.4} are given by 
\begin{align}
\label{denst}
w^{(q)}_T(r,Q^2) &=
\int^1_0 \ud \alpha \,\sum_{\lambda,\lambda'}
\left|
\psi^{(q)T}_{\lambda\lambda'}(\alpha,\mbf{r},Q) \right|^2 
\,,
\\
\label{densl}
w^{(q)}_L(r,Q^2) &=
\int^1_0 \ud \alpha\,\sum_{\lambda,\lambda'}
\left| 
\psi^{(q)L}_{\lambda\lambda'}(\alpha,\mbf{r},Q)\right|^2 \,,
\end{align}
where the squared photon wave functions (summed over quark helicities 
$\lambda$, $\lambda'$) are 
\be
\label{sumpsi+dens}
\sum_{\lambda, \lambda'} \left| \psi_{\lambda \lambda'}^{(q) T} (\alpha, \mbf{r},Q) \right|^2 
= 
\frac{3}{2 \pi^2} \, \alpha_{\rm em} Q_q^2 
\left\{ \left[ \alpha^2 + (1-\alpha)^2 \right] 
\epsilon_q^2 [K_1(\epsilon_q r) ]^2 
+ m_q^2 [K_0(\epsilon_q r) ]^2 
\right\} 
\ee
and 
\be
\label{sumpsiLdens}
\sum_{\lambda, \lambda'}  \left|\psi_{\lambda \lambda'}^{(q) L}(\alpha, \mbf{r},Q) \right|^2 
=
\frac{6}{\pi^2} \, \alpha_{\rm em} Q_q^2 
Q^2 [\alpha (1-\alpha)]^2 [K_0(\epsilon_q r) ]^2 
\ee
for transversely and longitudinally polarized photons, respectively. 
Here $r=\sqrt{\mbf{r}^2}$ with $\mbf{r}$ denoting the two-dimensional 
transverse vector from the antiquark to the quark. 
$Q_q$ are the quark charges in units of the proton charge, 
and $K_0$ and $K_1$ are modified Bessel functions. 
The quantity 
$\epsilon_q = \sqrt{\alpha (1-\alpha) Q^2 +m_q^2}$ involves the 
quark mass $m_q$. 
For a derivation of the photon wave functions 
see for example \cite{Ewerz:2006vd}.

\end{appendix}

\end{document}